\documentclass[prx,twocolumn,superscriptaddress]{revtex4-1}
\usepackage{amsmath}
\usepackage{amsfonts}
\usepackage{amssymb}
\usepackage{epsfig}
\usepackage{graphicx}
\usepackage{color}

\newcommand{\be}{\begin{equation}}
\newcommand{\ee}{\end{equation}}

\begin{document}

\title{Nonequilibrium equation of state in suspensions of active colloids}

\author{F\'elix Ginot}
\affiliation{Institut Lumi\`ere Mati\`ere, CNRS UMR 5306, 
Universit\'e Claude Bernard Lyon1, Universit\'e de Lyon, France}

\author{Isaac Theurkauff}
\affiliation{Institut Lumi\`ere Mati\`ere, CNRS UMR 5306, 
Universit\'e Claude Bernard Lyon1, Universit\'e de Lyon, France}

\author{Demian Levis}
\affiliation{Laboratoire Charles Coulomb, CNRS UMR 5221,  Universit\'e
Montpellier 2, Montpellier, France}

\author{Christophe Ybert}
\affiliation{Institut Lumi\`ere Mati\`ere, CNRS UMR 5306, 
Universit\'e Claude Bernard Lyon1, Universit\'e de Lyon, France}

\author{Lyd\'eric Bocquet}
\affiliation{Institut Lumi\`ere Mati\`ere, CNRS UMR 5306, 
Universit\'e Claude Bernard Lyon1, Universit\'e de Lyon, France}

\author{Ludovic Berthier}
\affiliation{Laboratoire Charles Coulomb, CNRS UMR 5221,  Universit\'e
Montpellier 2, Montpellier, France}

\author{C\'ecile Cottin-Bizonne}
\affiliation{Institut Lumi\`ere Mati\`ere, CNRS UMR 5306, 
Universit\'e Claude Bernard Lyon1, Universit\'e de Lyon, France}

\date{\today}

\begin{abstract}
Active colloids constitute a novel class of materials 
composed of colloidal-scale particles locally converting chemical energy 
into motility, mimicking micro-organisms. Evolving far from equilibrium, these
systems display structural organizations and dynamical properties 
distinct from thermalized colloidal 
assemblies. Harvesting the potential of this new 
class of systems requires the development of a conceptual framework 
to describe these intrinsically nonequilibrium systems. 
We use sedimentation experiments to probe the nonequilibrium 
equation of state of a bidimensional assembly of active 
Janus microspheres, and conduct computer simulations 
of a model of self-propelled hard disks. Self-propulsion 
profoundly affects the equation of state, but these changes 
can be rationalized using equilibrium concepts.
We show that active colloids behave, in the dilute limit, as an ideal 
gas with an activity-dependent effective temperature. 
At finite density, increasing
the activity is similar to increasing adhesion between 
equilibrium particles. We quantify this effective adhesion 
and obtain a unique scaling law relating activity and effective adhesion 
in both experiments and simulations. Our results provide a new and 
efficient way to understand the emergence of novel phases of matter 
in active colloidal suspensions. 
\end{abstract}

\maketitle

\section{Introduction}

Because they can be viewed as large atoms, colloids
have been used to perform detailed explorations of equilibrium 
materials. Experimental studies of colloidal systems 
have largely contributed to the development of 
accurate statistical mechanics descriptions of 
classical states of matter~\cite{Pusey86,Hansen2005}. 
Recently, `active colloidal matter'
has emerged as a novel class of colloidal systems, which is currently 
under intense scrutiny. Active colloids evolve 
far from equilibrium because they are self-propelled, motorized or 
motile objects, for which active forces compete with
interparticle interactions and thermal fluctuations.
This broad class of systems encompasses bacterial colonies, 
epithelial tissues and specifically engineered, or `synthetic', 
colloidal systems~\cite{Ball2013,Ramaswamy2010,Poon2013}. 
Progress in particle synthesis and tracking capability are such that 
frontier research in this field
has shifted from single particle studies to the understanding of 
physical properties of bulk active materials. Such 
large assemblies raise a number of basic 
physical challenges, from elucidation of two-body 
interactions to emerging many-body physics. In an effort to understand
the various phases of matter emerging in active colloids, we study
how self-propulsion affects the colloidal equation of state, and provide
a microscopic interpretation of our results. 

Experimentally, active colloidal systems have been 
mostly characterized so far through detailed single particle 
studies~\cite{Paxton2005,Howse2007} in the dilute gas limit, 
where particles only interact with the solvent. In this regime, 
a mapping from nonequilibrium active systems to 
equilibrium passive ones was established in the presence 
of a gravitational field, where activity renormalizes the value of 
the effective temperature~\cite{Tailleur2009,Palacci2010}.
Theoretical analysis shows that this experimental result
is not trivial, as the existence of a mapping to an effective 
`hot' ideal gas might break down for different 
geometries~\cite{Tailleur2009,Szamel2014} 
and depends also sensitively of the detailed modelling of
self-propulsion~\cite{Szamel2014,Encul}.

Much less is understood at finite densities
when particle interactions and many-body effects 
cannot be neglected. Several new phases of active matter 
have been observed experimentally in synthetic 
self-propelled colloids~\cite{Teurkauff2012,Palacci2013,Buttionni20013}, 
from active cluster phases at relatively low density to 
gel-like solids and phase separated systems at larger density,
suggesting that active colloids tend to aggregate when 
self-propulsion is increased. These observations have
triggered a number of theoretical studies~\cite{Ball2013b}, 
which have suggested several physical mechanisms for 
particle aggregation. In particular, particles may aggregate in regions 
of large density, because self-propulsion is hindered 
by steric effects~\cite{Tailleur2008,fily2012,redner2013,Levis2014}. 
Clustering is also observed when more complex
coupling between motility, density, reactant 
diffusion~\cite{Mognetti2013,stark2014,Encul}
are taken into account. Motility induced adhesion 
can be strong enough to induce a nonequilibrium phase separation
between a dense fluid and a dilute gas, as observed 
in some recent studies, but complete phase separation does not 
always occur in simulations~\cite{Levis2014} or 
experiments~\cite{Teurkauff2012,Palacci2013} on active systems.

Here, we characterize the phase behaviour and  equation 
of state of a system of active colloids 
and of a model of self-propelled hard disks, extending 
to nonequilibrium suspensions the classical sedimentation studies 
pioneered by Jean Perrin~\cite{perrin}. 
While theory and simulations have recently started to explore the equation 
of state of simple models of active particles~\cite{valeriani,manning,brady}, 
there is so far 
no corresponding experimental investigation of the pressure-density
relation in active particle systems. We find that 
activity modifies the equation of state in a way 
which can be described by the introduction of both an effective
temperature for the dilute system, and of an effective adhesive
interaction at finite density. From the known equation of state
for adhesive disks at equilibrium, we extract an effective 
adhesion between active particles and find a unique scaling law
relating activity to self-propulsion in both 
experiments and simulations.

\section{Sedimentation experiments in 
nonequilibrium active suspensions}

Sedimentation is a simple yet powerful tool to study 
colloidal suspensions, because it allows a continuous exploration of 
the phase behaviour of the system without fine-tuning of the 
volume fraction~\cite{perrin,piazzareview}. This technique
has been used to explore phase diagrams and equation of state
for several types of 
suspensions~\cite{Biben92,Biben93,piazza1,Chaikin96,piazza,durian}. 
Sedimentation allows us to extend  previous work on active colloids performed   
in the dilute regime~\cite{Palacci2010} and at 
low density~\cite{Teurkauff2012}, to a much broader range 
of densities.

\begin{figure}
\includegraphics[width=8.5cm]{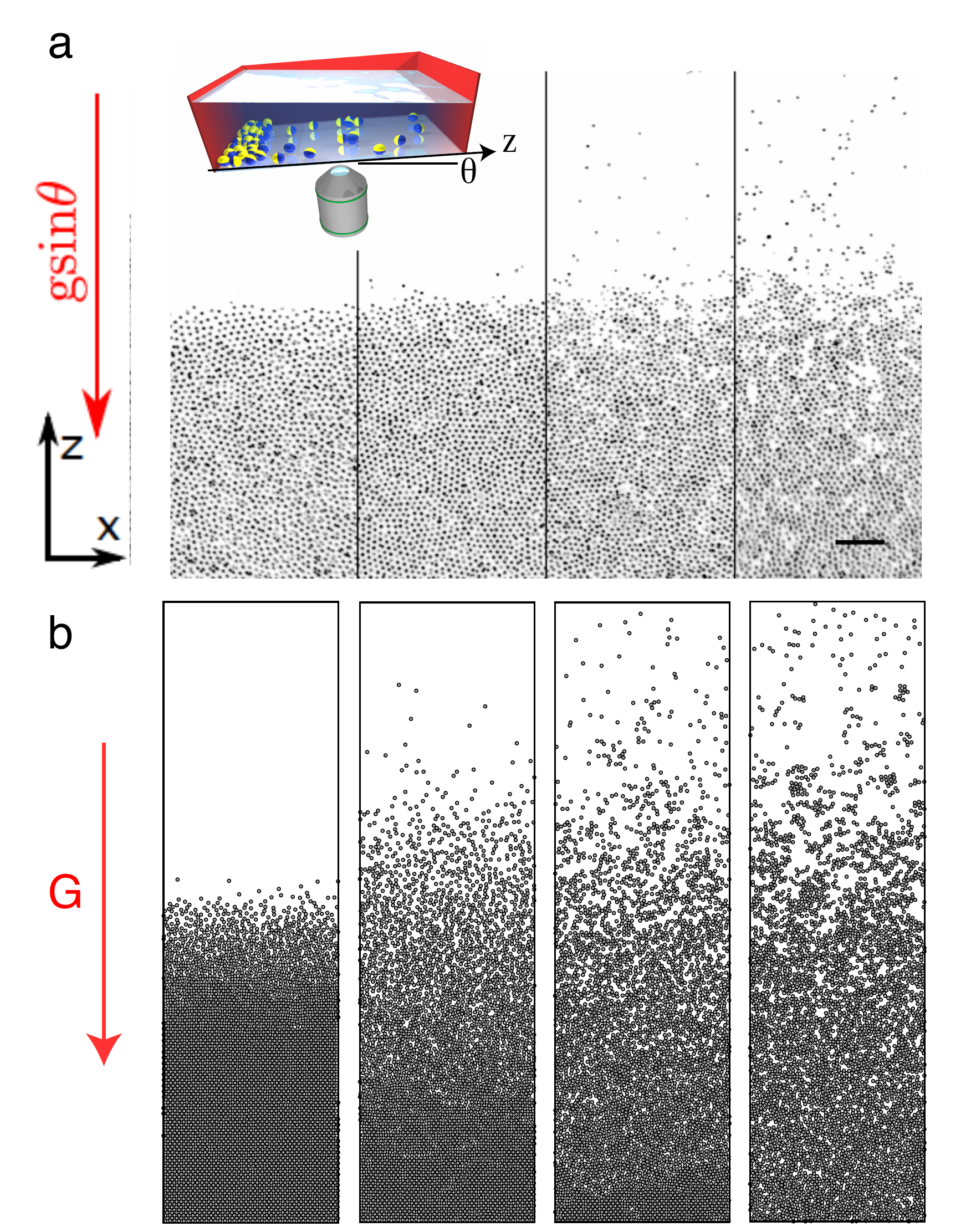}
\caption{{Imaging sedimentation profiles (color online).} 
Snapshots of (a) phoretic gold-platinum Janus colloids and (b)
self-propelled hard disks under gravity for activity increasing 
from left to right. 
(a) Four different activities are shown, 
corresponding to P\'eclet numbers $P_e = 0$ (passive case), 
8, 14, and 20 (from left to right). Inset shows a 
schematic view of the experimental setup.  The scale bar is 20~$\mu m$,
(b) Simulation snapshots for persistence times 
$\tau=0$ (passive disks), 1, 10, 100 (from left to right).}
 \label{sedim}
\end{figure}

Experimentally, we study colloidal particles which are self-propelled through 
phoretic effects. We use gold Janus 
microspheres with one half covered with platinum. When immersed 
in a hydrogen peroxide bath the colloids convert chemical energy 
into active motion. The average radius 
obtained from scanning electron microscopy
measurements (SEM) is $R=1.1\pm 0.1~\mu$m, but image analysis 
at large density indicates an effective radius between 1.34 and 
1.44~$\mu$m, decreasing slightly with activity. In the following, 
we assume that $R$ is constant, with the value given by the SEM measurements. 
Due to gravity, the Janus colloids (mass density $\rho \sim 11$~g.cm$^{-3}$) 
form a bidimensional monolayer at the bottom of the observation chamber.
We tilt the system 
by a small angle $\theta \sim 10^{-3}$~rad along the $z$-direction
as shown in the inset of Fig.~\ref{sedim}a, so that the gravity 
field felt by the particles is reduced to $g \sin \theta$. 
To study the effect of activity on the sedimentation, we measure the density
profiles in the $z$-direction for various mass concentrations in 
H$_2$O$_2$ from 0 to 10$^{-2} ~w/w$. This controls the level of particle 
self-propulsion, which can be quantified by the translational 
P\'eclet number $P_e$, defined as: $P_e=\frac{{Rv}}{{D_0}}$, 
where $D_0$ is the diffusion coefficient of the  
colloids in the absence of activity, and $v$ is the average velocity of 
free microswimmers (see Methods).  
Notice that the present active particle system was 
previously studied only at a single density~\cite{Teurkauff2012}, and 
the experimental setup to study its sedimentation profiles 
differs from earlier experimental studies~\cite{Palacci2010}.

We also conduct simulations of sedimentation 
profiles within a simple model of self-propelled 
hard disks~\cite{Berthier2013,Levis2014}, see Fig.~\ref{sedim}b.
The model is simpler than the experiments on two key aspects. Firstly, it
uses a purely hard core repulsion between particles. This is 
a reasonable choice, because the pair interaction between the Janus colloids 
has not been characterized in detail, and this avoids 
the introduction of multiple control parameters. Secondly, 
the phoretic mechanism behind self-propulsion is not simulated, 
and particle activity is implemented directly into Monte-Carlo equations of 
motion~\cite{Levis2014}, which generalize the algorithm 
used for Brownian hard disks to introduce self-propulsion. 
In the model, the activity is controlled by a single 
parameter, the persistence time $\tau$ (equivalent to a rotational 
P\'eclet number) determining the 
crossover time between ballistic and diffusive regimes in the dilute 
limit~\cite{Levis2014}. The intensity of the gravity force $G$ 
controls the sedimentation process (see Methods). 
The phase behavior and dynamics of the model without gravity has been 
studied carefully before~\cite{Berthier2013,Levis2014}, 
but the equation of state has never been analysed. 

In Fig.~\ref{sedim}, we show the sedimentation images for both the 
experimental system and the numerical model, for four different 
activity levels. The relevance of sedimentation studies 
is immediately clear, as the system continuously 
evolves from a very dilute suspension at the top, to very dense 
configurations at the bottom. Therefore, a single experiment
explores at once a broad range of densities for a given level of activity.
Despite its simplicity, the numerical model reproduces 
qualitatively the various regimes observed for Janus colloids.
For both systems, the passive case reveals equilibrium configurations from a
dilute fluid to a dense homogeneous amorphous phase, while active systems 
exhibit much richer structures. The dilute gas spreads 
over larger altitudes, finite size clusters are observed 
at moderate densities, gel-like configurations are found at larger densities, 
and dense, arrested phases exist at the bottom of the cell, as revealed
by these images. These various phases have been 
carefully analysed in the numerical model~\cite{Berthier2013,Levis2014},
but only the cluster phase was studied experimentally 
before~\cite{Teurkauff2012}. In the following we shall record 
and analyse quantitatively the equation of state of both systems over 
a range of densities where dynamics is not arrested and steady state 
conditions can be achieved.

\section{Sedimentation profiles: dilute 
limit and effective temperature}

\begin{figure*}
\includegraphics[width=17cm]{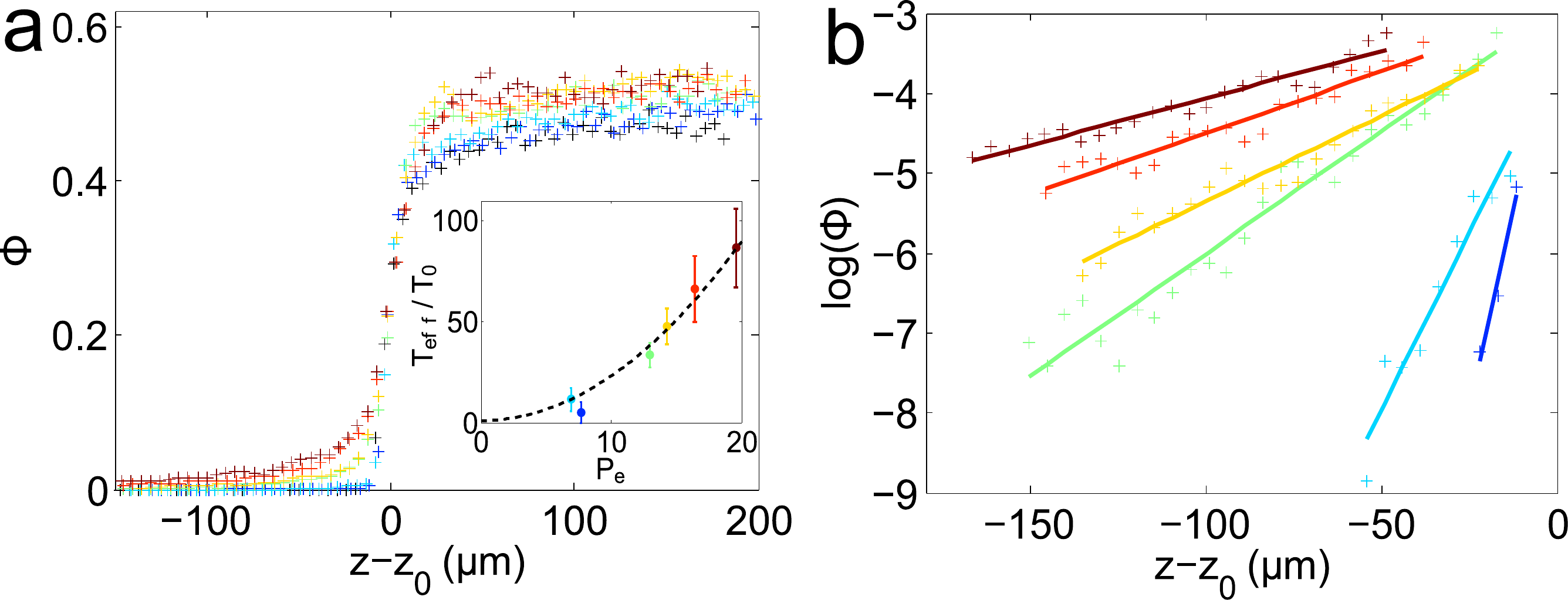}
\caption{{Experimental density profiles (color online)}. 
(a) Density profiles for various activities, arbitrarily 
shifted horizontally so that $z_0$ corresponds to $\phi = 0.2$.  
Inset: evolution of ${T}_{\textrm{eff}}$ as a function of the P\'eclet 
number, together with the theoretical expectation,
$T_{\rm eff}/T_0 = 1 + \frac{2}{9} Pe^2$.  
(b) Zoom on the dilute phase $\phi \ll 1$ using a semi-log scale, 
with the associated exponential fit, 
from which we extract the value of the effective temperature.}
\label{profile}
\end{figure*}

Our basic physical observable is the density profile $\phi(z)$
measured along the direction of the gravity from the images 
shown in Fig.~\ref{sedim} (see Methods).  
In Fig.~\ref{profile}a we show the density profiles obtained 
experimentally for different activities, which confirms the 
large range of densities explored in each experiment. 
We focus in Fig.~\ref{profile}b on the dilute phase
at small density (large $z$) and observe that for $\phi \lesssim 0.05$
the density evolves exponentially with the position
with a decay rate evolving continuously with the activity. We have 
obtained similar exponential profiles in the simulations. 

The experimental results suggest that in the dilute limit, 
the system behaves as an ideal gas with an activity-dependent 
effective temperature, ${T}_{\textrm{eff}}$, such that
$\phi(z) \sim \exp \left[ - {m g z \sin \theta}/(k_B 
T_{\rm eff}) \right]$ where $m$ is the mass of a colloid
and $k_B$ the Boltzman constant. 
From the linear fits shown in Fig.~\ref{profile}b, we extract the value 
of $k_B T_{\textrm{eff}} / (m g \sin \theta)$ for various activities. 
These measurements are more precise for $P_e$ above 10, 
as the dilute phase extends over larger distances;
$k_B T_0/(m g \sin \theta)$ is then evaluated from
the evolution of $k_B T_{\textrm{eff}} / (m g \sin \theta)$ with $P_e$.
Overall, our results in this regime agree 
with earlier work on dilute suspensions~\cite{Palacci2010}, 
and the theoretical expression, $T_{\rm eff} / T_0  = 1 + \frac{2}{9}{Pe}^2$  
is recovered, see inset of Fig.~\ref{profile}a. 
Quantitatively, we observe that $T_{\rm eff}$ increases 
from the thermal bath temperature $T_{\rm eff} \approx T_0 \approx 300$K 
for passive colloids  (from which we determine the tilt angle 
$\theta \approx 8.10^{-3} \pm 2.10^{-3}$) to a maximum value of 
about $T_{\rm eff} \approx 3.10^4$~K for the most active system.  

In the simulations, we find similarly that $\phi(z)$ 
decays exponentially with $z$. We have verified numerically that 
the dependence on the gravity field $G$ is given by 
$\phi(z) \sim \exp [ - z G / (k_B T_{\rm eff}) ]$, which 
defines an effective temperature.  
Exponential decay is obeyed over about 4 decades for  
$\phi \in [10^{-6}, 10^{-2}]$. The lower bound stems from 
statistical accuracy, but the upper bound emerges because profiles 
become non-exponential when density is large enough for many-body 
interactions to play a role. Numerically,
$T_{\rm eff}$ is directly proportional to the persistence 
time of the self-propulsion, $T_{\rm eff} \propto \tau$. 
Therefore, self-propelled hard disks under gravity also behave 
at low density, $\phi \ll 10^{-2}$, as an ideal gas with an 
effective temperature 
different from the bath temperature, $T_{\rm eff} > T_0$.

To reinforce this view, we performed additional simulations in the 
absence of gravity where we measured in the dilute limit $\phi \to 0$ both the 
self-diffusion coefficient $D_s$ (from the long-time limit of the mean-squared 
displacement) and the mobility $\mu$ (measured from the long-time limit of the 
response to a constant force in the linear response regime). 
The diffusion constant scales
with the persistence time $D_s \propto \tau$, as expected~\cite{Levis2014}, 
and we find that the mobility does not scale with $\tau$, such that 
$D_s / \mu \propto \tau$. 
Our simulations indicate that the equality 
\be 
k_B T_{\rm eff} = {D_s}/{\mu}
\label{einstein}
\ee
holds quantitatively, within our statistical accuracy. 
Equation (\ref{einstein}) states 
that the same effective temperature controls the sedimentation profiles
and the (effective) Einstein relation between diffusion and mobility. 
Hence, it is justified to describe self-propelled particles
as an `effective ideal gas'. These conclusions are far from trivial, as they
do not hold for all types of active particles
(run-and-tumble bacteria being a relevant 
counterexample~\cite{Tailleur2009}), while the mapping to an ideal gas 
may break down in more complex geometries even for  
self-propelled particles~\cite{Szamel2014}. 
Finally, note that $T_{\rm eff}$ is a single particle quantity, 
which is conceptually distinct from the (collective) effective 
temperature emerging in dense glassy regimes~\cite{BK2013}.   

\section{Nonequilibrium equation of state 
in active suspensions}

\begin{figure*}
\includegraphics[width=17cm]{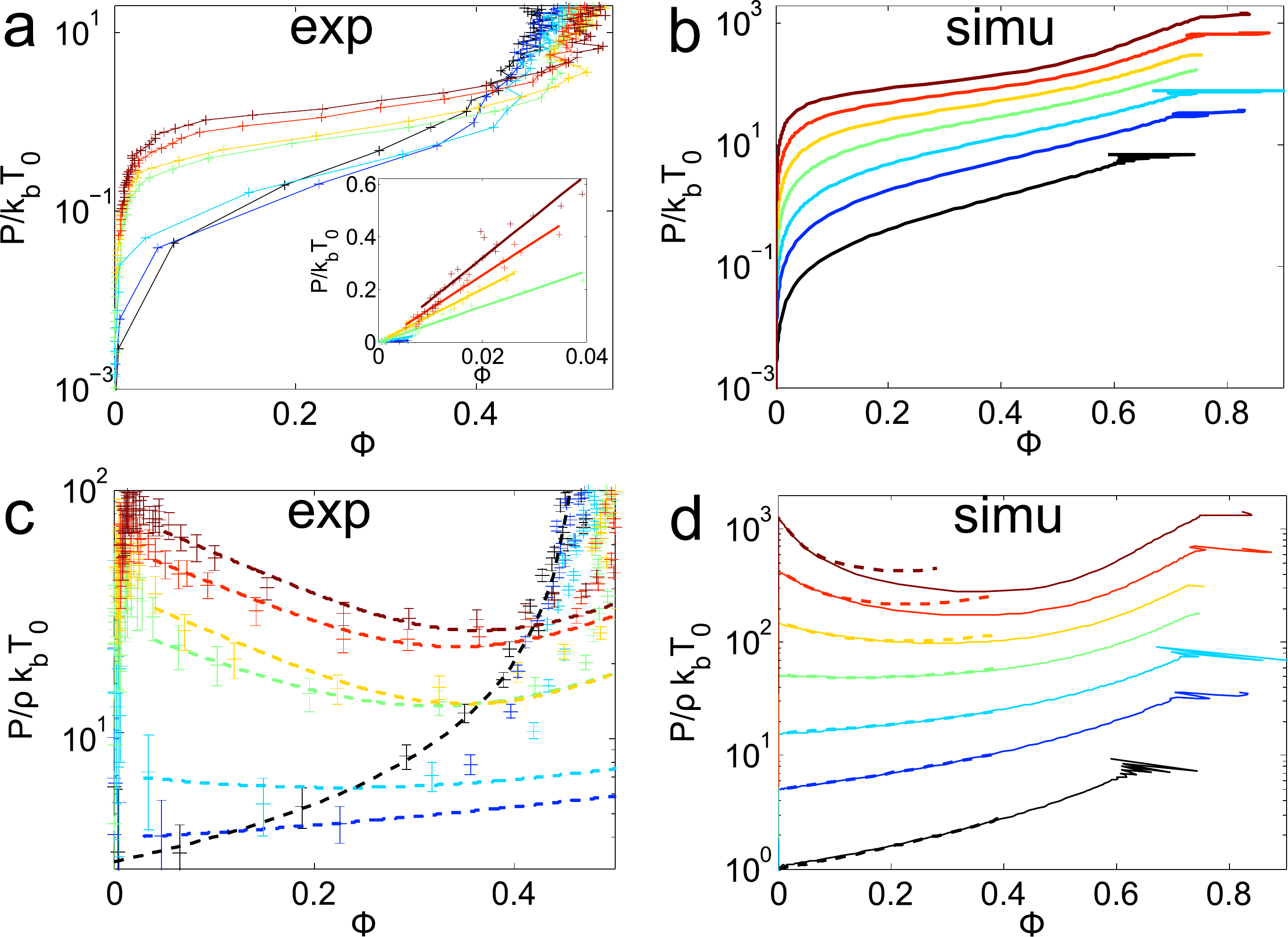}
\caption{{Equation of state and compressibility factor (color online)}. 
Equation of state $P/(k_B T_0)$ versus
linear density $\phi$ for experiment (a) and simulations (b).
In (a) the inset is a zoom on the dilute phase, where lines 
represent the best linear fit from which we recover $T_{\rm eff}$. 
(c) Compressiblity factor versus $\phi$ for experiment. Black points: 
passive case; black dashed curve: empirical equation of state for hard disks.
Dashed color lines obtained from Eq.~(\ref{viriel}).
(d) Compressiblity factor versus $\phi$ for simulations with 
dashed lines obtained from Percus-Yevick closure for two-dimensional
Baxter model.
Experiments: $T_{\textrm{eff}}/T_0= 1$, 5, 15, 34, 47, 62, 87 (bottom to top).
Simulations: $\tau = 0, 1, 3, 10, 30, 100, 300, 1000$ (bottom to top).}  
\label{pressionrho}
\end{figure*}

Provided we measure density profiles in steady state conditions, 
and using the sole assumption of mechanical equilibrium~\cite{Biben92}, 
we can convert the measured profiles 
into a  pressure measurement, 
$P(z) = \Pi(z)-\Pi_{0}$, where $\Pi_{0}$ is the pressure at the top of 
the observation cell and $ \Pi(z)= \frac{m g \sin \theta}{\pi R^2 } 
\int_{z}^{L} dz' {\phi}(z')$. We can then represent the parametric evolution of
$P(z)$ with $\phi(z)$ by varying $z$, which gives direct access 
to the nonequilibrium equation of state $P(\phi)$. 
To our knowledge, there is no previous experimental report of such 
quantities, that have only very recently been discussed 
in theoretical work~\cite{valeriani,manning,brady}. 
In Figs.~\ref{pressionrho}a,b
we present the outcome of this analysis, and show 
the evolution of the normalised osmotic pressure $P/({k}_{B}{T}_0)$ 
for various activities in experiments and simulations.
In Figs.~\ref{pressionrho}c,d we replot the same set of data 
to examine how the compressibility factor 
$Z = {P} / (\rho {k}_{B} {T}_{0})$, with $\rho$ the number density, 
depends on density $\phi$. This 
second representation offers a finer perspective 
on the nonequilibrium equation of state, as the ideal gas 
behaviour at low $\phi$ is scaled out. 

In the dilute regime, we recover the `effective' 
ideal gas behavior, $P = \rho k_B T_{\rm eff}$, which directly follows
from integration of the exponential profiles of 
Fig.~\ref{profile}. This linear dependence of $P$ with $\phi$  
is more carefully examined in Figs.~\ref{pressionrho}c,d, 
as it translates into a finite value for $Z(\phi \to 0)$, namely 
$Z(\phi) \to T_{\rm eff} / T_0$, as observed. The data in 
Figs.~\ref{pressionrho}c,d thus provide a simple, direct measurement 
of the effective temperature in active suspensions.

More interesting is the behaviour at finite density,
which has not been explored experimentally before. Both 
experimental and numerical data indicate that the functional form of  
the equation of state changes continuously 
as the activity increases, in a way that
cannot be uniquely accounted by the introduction of the effective 
temperature. This is not surprising, as the images in Fig.~\ref{sedim}
show that the structure of the system at finite density changes 
dramatically with activity.
To confirm this, we first analyse the data for passive systems. 
In that case, the equation of state can be well 
described by an empirical equation of state for hard 
disks~\cite{Santos1995,note}. The agreement with 
hard disks is not guaranteed in experiments, because it 
is not obvious that (passive) Janus colloids uniquely interact with 
hard core repulsion, but this seems to be a good approximation. 
In the simulation, Fig.~\ref{pressionrho}d (black curves), 
we show that the data for passive disks agree very well with the 
equilibrium Percus-Yevick equation of state. This is expected, because 
the system becomes for $\tau=0$ an equilibrium fluid of hard disks. 

When activity increases, experiments and simulations 
can no longer be described by the equilibrium hard disk equation of state. 
Qualitatively, the compressibility 
factor $Z(\phi)$ increases more weakly with $\phi$ for moderately 
active particles than for the passive system, and we observe that 
$Z(\phi)$ even decreases with $\phi$ for 
more active systems at low $\phi$, so that $Z(\phi)$ actually becomes 
a non-monotonic function of $\phi$ for large 
activity~\cite{valeriani,manning,brady}.
In equilibrium systems, such a behaviour represents 
a direct signature of adhesive interactions~\cite{piazza}. 
This suggests that the strong clustering observed in 
Fig.~\ref{sedim} in active particle systems directly impacts
the equation of state, which takes a form reminiscent of  
equilibrium colloidal systems with attractive interactions. 

\section{Determination of effective adhesion 
induced by self-propulsion} 

The idea that self-propulsion provides a mechanism for 
inducing effective attractive forces in purely repulsive 
systems of active particles has recently 
emerged~\cite{Buttionni20013,Tailleur2008,brady,epl2013}.   
Because we have direct access to the equation of state 
in our systems of active particles, we are in a unique position to 
study the similarity between the equation of state of nonequilibrium 
active particles and equilibrium adhesive disks. If successful, 
we can then quantify the strength of the effective adhesion 
between particles which is induced by the self-propulsion mechanism. 

To this end, we compare the equation of state 
obtained for active particles to the equilibrium equation 
of state of a system of adhesive particles. We have analysed 
the equation of state of the Baxter model of 
adhesive particles in two dimensions~\cite{baxter}.  
This is a square-well potential with a hard core repulsion 
for $r < \sigma$, and a short-range repulsion of range $\sigma+\delta$
and depth 
\be
V(\sigma < r < \sigma+\delta) = - k_B T \ln \left( 
\frac{\sigma+\delta}{4 \delta}  A \right),  
\ee
where $A$ is a nondimensional number quantifying the adhesion strength,
and $V(r > \sigma+\delta)=0$. 
The model is defined such that the adhesive limit can smoothly 
be taken, where $\delta \to 0$, $V(r=\sigma^+) \to \infty$ but the second 
Virial coefficient remains finite and is uniquely controlled by 
the dimensionless 
adhesion parameter $A$~\cite{baxter,noro}. The purely repulsive
hard disk limit is recovered with $A \to 0$.  
For this model, the first two Virial 
coefficients are known analytically~\cite{Post1986}:
\begin{equation}
Z=\frac{{P}}{\rho {k}_{B} {T}}=1+b_1\phi+b_2\phi^2+{\cal O}(\phi^3),
\label{viriel}
\end{equation}
with $b_1=2-A$, $b_2=\frac{25}{8}-\frac{25}{8}A+\frac{4}{3}A^2-0.122A^3$.
When $A$ increases, the compressibility factor $Z$ in Eq.~(\ref{viriel})
changes from the monotonic hard disk behaviour to a non-monotonic 
density dependence for $A>2$ as the initial slope given by $b_1$
then becomes negative. 

\begin{figure}
\includegraphics[width=8.5cm]{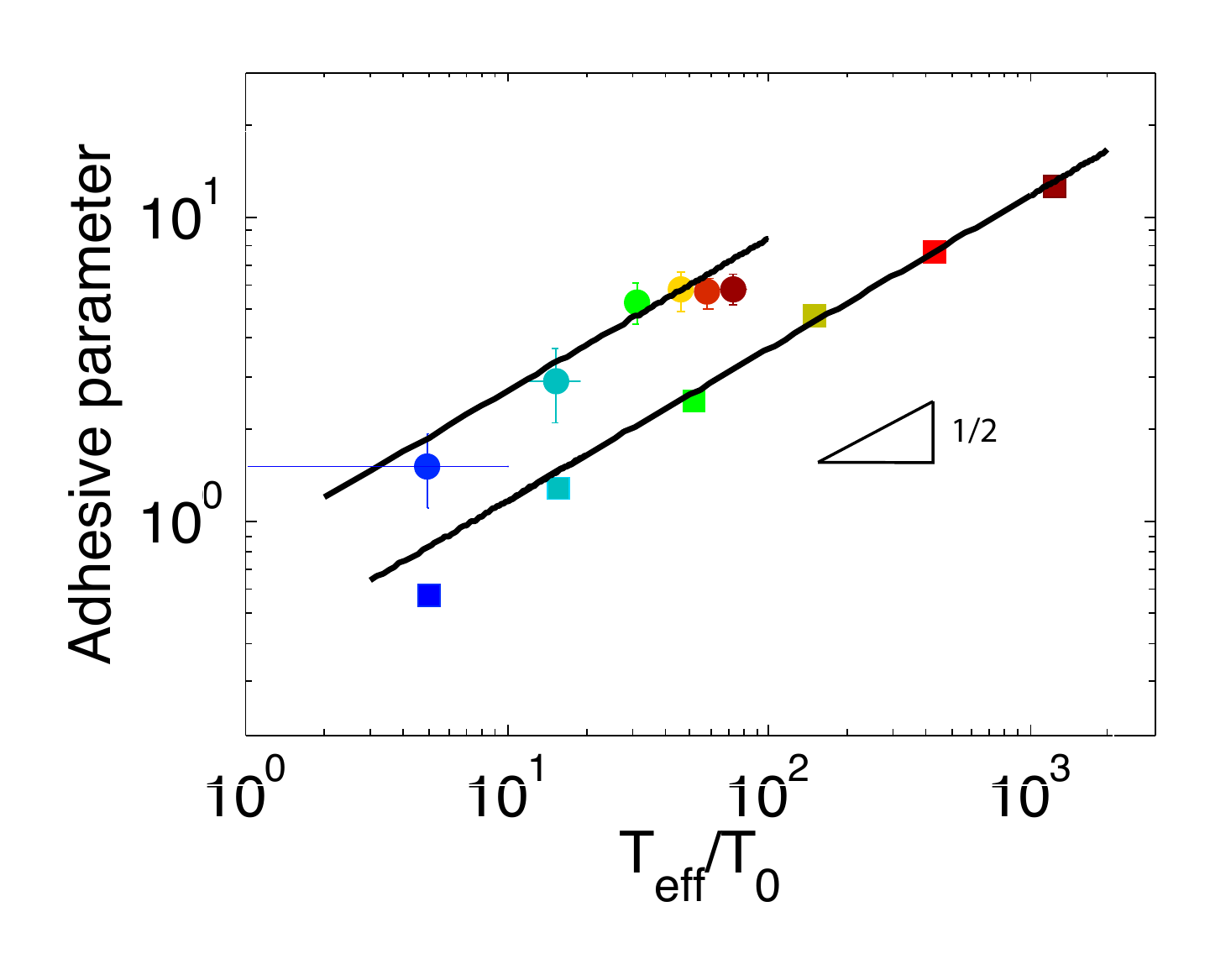}
\caption{{Quantifying motility-induced adhesion (color online)}. 
Evolution of adhesive parameter $A$ obtained from the equation of 
state with the effective temperature obtained in the dilute limit
for both experiments (circles) and simulations (squares).
The lines correspond to the scaling law $A \sim \sqrt{T_{\rm eff}/T_0}$,
with slighlty different prefactors for simulations and experiments.} 
\label{ATeff}
\end{figure}

The above expression provides a reasonable 
description of the experimental data, provided we carefully 
adjust $A$ for each activity. This 
provides a direct estimate of the `effective adhesion' 
induced by the self-propulsion, which thus quantifies 
the nonequilibrium clustering of self-propelled particles 
using a concept drawn from equilibrium physics. In practice, 
we fit the experimental compressibility factor at intermediate densities 
($0.04<\phi<0.4$) with the expression $Z = \alpha \left( 1+b_1\phi+b_2\phi^2
+{\cal O}(\phi^3) \right)$, adjusting both the effective adhesion $A$ and 
a prefactor $\alpha$, which accounts for the effective temperature.
Within errorbars, we obtain $\alpha \approx T_{\rm eff}/T_0$, 
confirming the robustness of the analysis. 
We have also verified that the uncertainty on 
the determination of $\phi$ (due to the uncertainty in 
particle diameter) has a negligible impact on the measured $A$ values. 
As shown in Fig.~\ref{ATeff}, we find that the 
effective adhesion $A$ increases with self-propulsion. 
 
We carry out a similar analysis for the compressibility factors
obtained numerically for self-propelled hard disks. Because the 
numerical data are statistically accurate over a broad density range, 
we solved the full Percus-Yevick closure relation to obtain the equation 
of state of the bidimensional Baxter model, 
and adjust $A$ to obtain the best agreement between
self-propelled and adhesive disks, as shown in Fig.~\ref{pressionrho}d.
As observed in the experiments, we again obtain an adhesion
$A$ that increases with self-propulsion, see Fig.~\ref{ATeff}. 
 
To analyse the evolution of the effective adhesion with activity, 
it is useful to represent $A$ as a function of $T_{\rm eff}/T_0$. 
This representation is convenient because it only involves
dimensionless quantities, and $T_{\rm eff}/T_0$ 
efficiently quantifies self-propulsion with no reference to 
its detailed physical origin ($T_{\rm eff}$ is
proportional to the diffusion constant of noninteracting particles).
Remarkably, we observe a similar scaling relation 
for both experiments and simulations between 
adhesion and effective temperature, namely $A \sim \sqrt{T_{\rm eff}/T_0} $.  

\section{Physical discussion}

The scaling law relating $A$ and $T_{\rm eff}$ suggests that
self-propelled disks at finite density do not simply
behave as a `hot' suspension of hard disks, but rather as a
hot suspension of adhesive disks with a simple
relation between adhesion and effective temperature. 
A plausible scaling argument for the emergence of such a scaling 
is to compare the 
contact duration for active particles, controlled 
by the rotational P\'eclet number (or persistence time), 
to a bonding time in the context of the Baxter model, 
$t_{\rm bond} \sim  \exp(-V/k_BT) \sim A$. This simple 
argument suggests however an incorrect proportionality
between $A$ and $T_{\rm eff}$. This demonstrates that the scaling 
of the active adhesion follows from a more complex interplay 
between bonding time and structural evolution induced by the self-propulsion.

To rationalize the observed scaling, 
we have compared quantitatively the structure 
obtained with self-propelled disks at finite density to 
computer simulations of the equilibrium Baxter model of adhesive 
disks~\cite{Post1986}. In particular, we find that 
the degree of clustering (quantified by the average cluster size)
is the same in both models when $A$ and $\sqrt{T_{\rm eff}}$ increase 
proportionally, which gives further support to the 
results found in Fig.~\ref{ATeff}. In previous work, 
it was shown that the average cluster size increases 
as $\sqrt{T_{\rm eff}}$ in our numerical model~\cite{Levis2014}. 
This $\sqrt{T_{\rm eff}}$ scaling law is actually captured by a kinetic model 
of reversible aggregation~\cite{Levis2014}, which assumes 
that clusters result from a competition between particle 
aggregation (when a self-propelled particle collides with an existing cluster)
and escape from the cluster surface (when the direction of the 
self-propulsion for a surface particle becomes oriented towards the exterior). 
While this physical model is exactly in the spirit as the above 
scaling argument based on a bonding timescale, the kinetic model 
is able to capture the many-body physics 
of particle exchanges between dynamic clusters responsible for the observed 
connection between adhesion and activity.

The results in Fig.~\ref{ATeff} indicate that, while simulations and 
experiments obey the same scaling,
the experimental adhesive parameter is systematically larger.
This suggests that a second mechanism could be at work in 
experiments. Following Refs.~\cite{Teurkauff2012,stark2014} 
we may attribute it to the chemical (phoretic) interactions between colloids. 
Since the fuel powering phoretic motion is depleted around the colloids, 
this may induce a diffusio-phoretic 
attraction between colloids.
From the force balance between colloids, this interaction can be quantified 
in terms of an attraction energy which is 
directly proportional to the chemical consumption rate, and therefore 
to the self-propelled velocity~\cite{Teurkauff2012}. 
Altogether, this suggests an attraction energy 
scaling like $A \sim P_e \sim \sqrt{T_{\rm eff}}$, as observed.
Our experiments and simulations therefore point to an
additive combination of kinetic and phoretic contributions to the 
experimentally observed effective attraction. 

Our approach provides a quantitative way to account for
the clustering commonly observed in self-propelled particle 
systems as it suggests that active particles effectively 
behave as an equilibrium system of hot adhesive particles, 
at least in the regime explored experimentally in this work.  
In the experiments, we can 
notice that the effective adhesion $A$ seems to saturate with the 
effective temperature, see Fig.~\ref{ATeff}. This could be linked 
to the fact that in our experiments we do not observe 
a macroscopic phase separation. In adhesive spheres, phase separation 
occurs for larger adhesion values, $A > 10$~\cite{baxter},
while clustering is observed~\cite{Post1986} 
for adhesion values corresponding to our experiments. 
Finally, it is remarkable that 
despite the complex mechanisms responsible for 
self-propulsion at the micro-scale, the experimental equation of state 
is very well reproduced in a very simple model of self-propelled 
hard disks.
This suggests that Janus active colloids represent an interesting 
experimental model system to study 
the statistical physics of `active colloidal matter'. 

\section*{ACKOWLEDGMENTS}
F. Ginot and I. Theurkauff contributed equally to this work.
We thank T. Biben, H. L\"owen, and G. Szamel 
for fruitful discussions, B. Ab\'ecassis for the 
synthesis of the gold colloids and INL for access to its clean room facility,
A. Ikeda for his help in the numerical integration of the Percus-Yevick 
equation.  
The research leading to these results has received funding
from the European Research Council under the European Union's Seventh
Framework Programme (FP7/2007-2013) / ERC Grant agreement No 306845.\\

\appendix

\section{Experiments on Janus active colloids.} 

We typically study the sedimentation of  about $10^6$ Janus colloids. After 
each increase in peroxide, we wait for 5
minutes to reach a stationary state inside the cell. The chamber depth 
is about 6~mm for a total volume in H$_2$O$_2$ of 300~$\mu$L so that there 
is sufficient fuel for the whole duration of the experiment (see movie in 
Supplementary Materials). In those conditions particles remain active 
during several hours.
The Janus colloids behave as small batteries, disproportionating 
the peroxide into water and dissolved dioxygen \cite{Posner2011}. These 
chemical reactions create electrochemical gradients around the colloid 
that lead to motion 
at $v \sim 2\mu{\rm m}.{\rm s}^{-1}$, depending on the concentration in 
peroxide. The 
P\'eclet number  can be used to describe the activity of the particles and 
is defined as: $P_e=\frac{{Rv}}{{D}}$, where $D=0.21\pm 0.05 
\mu\textrm{m}^2 . \textrm{s}^{-1}$ is the diffusion coefficient of the  
colloids in the absence of activity, and $v$ is the average velocity of 
free microswimmers, defined as colloids without neighbors closer than 6~$\mu$m.
Our basic observable for the analysis  is the density 
profile $\phi(z)$.  In the experimental system, we perform a very basic 
coarse-graining operation on pictures, counting the number of colloids 
${n}(x,z)$ in square boxes of size $l \sim 5\mu {\rm m}$. Then we compute 
the number density: $\rho(z)=\frac{1}{{L}}\intop_{0}^{{L}}\frac{n(x,z)}{l^2} 
dx$ and the 
the density profile  $\phi(z)= 
\frac{\pi {R}^2}{{L}}\intop_{0}^{{L}}\frac{{n}(x,z)}{l^2}dx=\pi {R}^2\rho(z)$ 
where $L \times L=350 
\times 350 \mu$m$^2$ is the size of the observation area.
For each peroxide concentration, 2000 4Mpx frames are taken at 20Hz using a 
Baumer HXC40 camera mounted on a Leica DMI 4000B microscope, and a custom 
made external darkfield ligthning ring. Particles trajectories are 
reconstructed using ordinary tracking algorithms. The error bars  are 
determined from the standard deviation of the fluctuations of independent 
measurements of the density obtained by coarse graining the sample in 60 
independent boxes along the $x$ axis. Moreover, as we are in a stationary 
state, this statistics was completed by 10 independent realizations. \\

\section{Simulations of self-propelled hard disks} 

To study sedimentation profiles, we introduce a hard wall at position
$z=0$, while keeping periodic boundary conditions in the $x$ direction. 
We introduce gravity as a force $\vec{G}$ acting along the $z$ direction
(see Fig. \ref{sedim}b), so that a particle displacement $\Delta z$ along 
the $z$ direction is accepted with the Metropolis acceptance rule
$\min [ 1, \exp( - \beta \Delta z G) ]$, where $\beta = (k_B T)^{-1}$.  
Under the influence of gravity, particles accumulate at the bottom 
of the simulation cell, whose height is taken large enough that 
the top wall does not influence the results. We report results for 
simulations with $N=6144$ disks in a box of width $W = 40 \sigma$, where
$\sigma$ is the disk diameter. We have performed tests with 
$N=1024$, 3072, and 12288 disks to establish that the present results 
are not influenced by finite size effects. 
For each value of $\tau$ controlling
the self-propulsion, we performed simulations for several values 
of the gravity field $G$, and checked that the results 
presented below for the equations of state obtained in steady state 
conditions do not depend on the chosen value. The density profile  
 $\phi(z)$  is obtained by performing very long simulations 
of the steady state profiles, and performing a time average for each 
pair $(\tau, G)$.  \\

\end{document}